# Two factor authentication using EEG augmented passwords

Ivan Švogor, Tonimir Kišasondi

*Faculty of organization and informatics, University of Zagreb, Pavlinska 2, 42000 Varaždin*
*E-mail(s): isvogor@foi.hr, tkisason@foi.hr*

**Abstract**. *The current research with EEG devices in the user authentication context has some deficiencies that address expensive equipment, the requirement of laboratory conditions and applicability. In this research we address this issue by using widely available and inexpensive EEG device to verify its capability for authentication. As a part of this research, we developed two phase authentication that enables users to enhance their password with the mental state by breaking the password into smaller, marry them with mental state, and generate one time pad for a secure session.*

**Keywords.** EEG, BCI, password augmentation, security, brain waves

## 1. Introduction

The current momentum of information age leads us to application of technology in almost every segment of daily life. Computer systems must handle more users; hence they need to be more powerful, robust and secure. Every new user is a potential threat, not because he will intentionally try to harm the security of a system, but because his carelessness may compromise the systems security. Such carelessness is apparent while entering a PIN, choosing a simple insecure password, or becoming a victim to phishing and other forms of social engineering. Despite the significant efforts in the field of information security to eliminate those problems, they continue to exist and form a continuous challenge.

Since the password is currently one of the generally accepted measures for secure authentication, we will concentrate our effort to strengthening it. The current research of this area is oriented towards determining security requirements for passwords along with combining it with some biometric characteristic of the user. The selected method largely depends on the desired security level. In this paper we will consider specialized systems using passwords, strengthening it to prevent reply attack and also demanding a certain mental state of the user entering it. This will be done with a modified one time password algorithm and the application of typical EEG device available on the market.

## 2. Current research

Current research in this area can be categorized in the papers engaged with theoretical or laboratory application of EEG authentication and two-phase EEG authentication. Major research efforts in recent years were directed towards application of brain waves as a biometric characteristic that is unique and inherent in every person. The main challenge of this research is recreating the brainwave. EEG as a biometric characteristic lacks constancy which depends on stress, fatigue, medication, environment (electrical equipment) etc. To cope with this, researchers often use some kind of stimuli to help in recreating the valid authentication EEG pattern [5], e.g. visualization of 3D object manipulation, counting, imagining letters and texts [4] etc. The authors [1] [4] used a visual stimuli, in an attempt to recreate the authentication pattern of a brain wave. They use one classifier per user to identify him/her which has proven to work properly. Alongside visual stimuli, audio stimuli can also be applied. In [3] authors use auditory evoked responses in a head authentication technique. The method used here proved to be very effective with a potential in telephony, since it uses voice and brain waves recorded at the ears. With EEG one can also determine the physical movement of a person. In [2] authors use this approach and identify for steps in an authentication procedure of the user. Direct application of brain wave patterns as a feature to authenticate a user is applied in [6] where the authors use pass-thought, a pattern of brain activity that acts as a password. While there are good theoretical advices of the research direction, there are some ethical questions that address the application of brainwaves for user identification and authentication. Another interesting research direction uses brainwaves (EEG signals) as a supplement to other methods [7]. The issue of current research is the applicability if the developed systems in real word environment, since the used equipment is expensive and the experiments are done in special laboratory environment. Another issue is the way that this characteristic is extracted from the user. With this in mind it is not realistic to expect great popularity in everyday use, but it is realistic to expect its application in some special cases. When sensors for portable EEG devices become cheaper and better, there is a good possibility of its application like [3] suggests. Most of users still prefer passwords since they are simple and relatively safe if used correctly. The greatest threats for passwords are weak, easy to guess passwords, password reuse and shoulder surfing. The work [8] enumerates 16 current challenges that passwords face today. Hence there are obvious shortcomings of current research in this area that is related to expensive equipment and laboratory conditions, in this paper we propose several research goals; establish the applicability of a typical EEG device currently available on the market, develop a technique that strengthens passwords with extra information that determines users mental state, and finally use EEG waves in a two phase authentication to increase the security of the password with a method that is a biometrically enhanced password which we call *EEGPass*.

## 3. Our approach

For the preparation of this research we made a test using a MindWave EEG device. This device is widely available on the market and it has a low price. Specification claims the ability to measure alpha, beta, gamma and delta waves, with the identification of relaxation, concentration, and blink detection. To verify these claims, we conducted an initial study, which involved six test subjects. For each test subject, we measured EEG signal and compared with successive measurements. The subjects had an alphabet displayed in front of them, and their task was to read letters in any way they like (but to repeat this way in subsequent measurements). These results showed no significant correlation between subjects or in successive measurements for the same subject. This is contrary with previous studies. We claim it to be because our measurements were significantly shorter (in [2] recordings were 20 minutes long), non-laboratory conditions, and significantly lower equipment. We had those conditions intentionally because users cannot use the system in laboratory conditions, with expensive equipment and wait for 20 minutes to authenticate. This led us to conclude that MindWive EEG will not be a good choice for a system based on pass-thought. Where MindWave proved effective is the recognising the mental state of the user, i.e. using alpha and beta waves to determine users' relaxation and concentration.

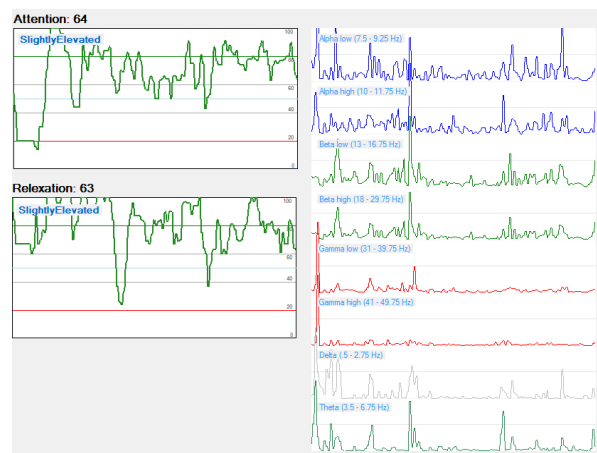

**Figure1. Sample measurement from preliminary test**

As we already said, MindWave cannot be used for a pass-thought based system but the EEG signals that it provides can be used as additional parameters for a password. Therefore we need to create an algorithm that will take advantage of the mental state of users (relaxation and concentration) to further enhance the password. The password will be divided into shorter segments, and for each segment, the user will select the mental state which must be matched to a segment in order for the full password to be accepted.

## 4. Pels and EEG signal state

### 4.1. Design of the solution

To use our EEGPass algorithm we need to set some initial data. If we just add the value of the each brain wave for relaxation and concentration to each character in the password, we can add additional 25 states from our EEG per character. Since that would prove additional benefits in terms of password cracking resistance, but would hardly make sense in an economical way and would not provide true two factor authentication. Since EEG based two factor authentication makes sense in high security areas, we wanted to tie a specific workstation (EEG unit) to a centralized authentication service. Unfortunately with MindWave devices that is not possible because we can only obtain the bluetooth MAC address, which is not advisable to use as it can be easily spoofed. Our approach is to use a variant of the hashed message authentication code since we want the client to be authenticated to a service. This decision was made because if we want to obtain brain wave signals; we must have a software client on the host machine to interface with the MindWave device. Also, we assessed the problem of malware on the host machine. Unfortunately, there are no real practical methods to protect any host software implementation from malware and advanced rootkits.

To augment passwords with EEG signals, we divide the password into shorter segments, called password elements, or *pels*. Each pel is further assigned the values of the EEG signal i.e. concentration and/or relaxation. Those parameters are scaled to 5 levels:

- low activity: S
- reduced activity: L
- normal activity: N
- increased activity: R
- high-activity: H
- Irrelevant: any activity is accepted: 0

For each character in the password, we will assign the required attention and relaxation brain wave pattern. So for example the password $qwerty123$ would be written as the following array with the following traits:

$[[q, H, 0], [w, H, 0], [e, H, 0], [r, 0, H], [t, 0, H], [y, 0, H], [1, H, 0], [2, H, 0], [3, H, 0]]$

Characters *qwe* are written with high attention, with irrelevant relaxation.
Characters *rty* are written with high relaxation, with irrelevant attention.
Characters *123* are written with high attention, with irrelevant relaxation.

In our current implementation, the PEL is equal to the span of characters that have the same required brain wave pattern. This password transformed into a pel array with 3 pels will be:

$[[qwe, H, 0], [rty, 0, H], [123, H, 0]]$

for example, lets name them $p_1$, $p_2$ and $p_3$. We grouped the same letters into pels because we want to be able to recognize any order the letters are entered. That way, the password can be entered as any combination of $p_1$ to $p_3$, for example $p_1 p_2 p_3$ is as valid as $p_3 p_2 p_1$ or $p_2 p_3 p_1$ etc. This adds a additional security measure if the user is employing bad password management and is reusing the password elsewhere, he can enter it in a different order just to avoid shoulder surfing attacks if there are other persons near him.

The EEG signals can be easily affected and governed by the user but they cannot be easily reproduced under conditions of stress, anxiety, drowsiness, effects of alcohol or drugs, etc. This way, the user has some control on the condition in which he can authenticate. Furthermore, while using our system the user can monitor the level of their concentration and relaxation. At a time when one of the two signals coincides a wanted pel the user can enter their pel. Along with previously described mental state that user can choose, this algorithm also increases the password resistance against password cracking attacks, since for each PEL which can be a single letter in the password we add two additional variables: relaxation and attention. Our preliminary research has shown that the process of adding different brain wave patterns other than attention alpha and relaxation beta seems to be problematic because other brainwave patterns seem to be harder to induce in some people, especially in environments requiring fast authentication, with lower budget and less invasive equipment.

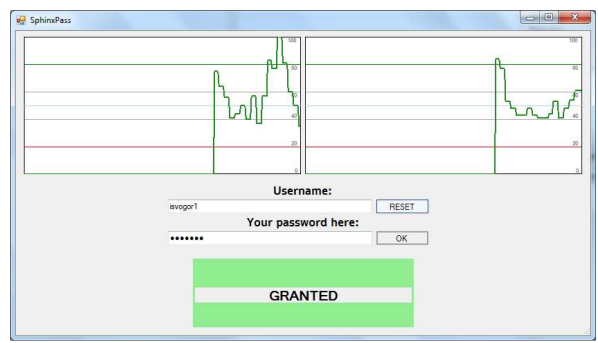

**Figure 2. Successful user login**

Since we are building a two factor authentication system which uses biometric components, we need to have a client based part of our application which interfaces with the hardware. Therefore, we can build a client server type model where the client part can interface with common protocols or methods like RADIUS or PAM or with custom server based apps. The server part of the app can be dislocated or combined on the same machine. Since this infrastructure is mandatory, we decided to strengthen the process with additional cryptography which will be basis for our authentication protocol.

### 4.2. Client part

The EEGPass has the following requirements:

1. Each client has a unique shared secret key that is known only by the server and client. This key will be our secret key for the HMAC (hash-based message authentication code as per RFC2104 / FIPS-PUB 198 [11, 12] with SHA-256 or better). We shall reference this key as $K$
2. The server needs to store $HP$ for pel $P$ such that $HP = HMAC(K, P)$ for each pel.
3. The system administrator while installing the client software and hardware enters the client's secret key $K$ in the client and server.

Enrolment phase:

1. Enrolment can be done on the client or server, but we suggest the enrolment on server so that the client doesn't need to transfer the enrolment data to server (where we skip step 5)
2. The user chooses his password and mental states
3. The user enrols into the system multiple times so that the client part of the application can be group the password into pels
4. For pels $p_1$ to $p_n$ we calculate $HP_1$ to $HP_n$
5. Client sends the server $HP_1$ to $HP_n$
6. Server stores the $HP_1$ to $HP_n$
7. Server calculates $HPF$ where: $HPF = HMAC(K, (HP1 + HP2+ … +HPn))$ where the + sign signifies string concatenation. $HPF$'s are calculated for all permutations of $HP_1$ to $HP_n$ which he has stored. $HP_1$ to $HP_n$ can be dropped, only $HPF_1$ to $HPF_n$ are needed for authentication.
8. System requires verification of the enrolment. The user authenticates 2 times with normal linear $P_1, P_2, …, P_n$ pels and two times with user specified randomized pels.

The authentication is performed in the following steps:

Client side:

1. User enters his pels form $P_1$ to $P_n$ and the order is not relevant.
2. Client calculates $HP_1$ to $HP_n$ and calculates the final $HPF$ as $HPF = HMAC(K, (HP1 + HP2+ … +HPn))$ where the + sign signifies string concatenation.
3. HPF is the final hashed pel which can be used as an equivalent of a hashed password

Server side:

1. Server receives $HPF$
2. If the received $HPF$ is in the pool of $HPF$'s on the server, the authentication is successful.

It is possible to extend this algorithm to work as a onetime passwords base algorithm. To do that, we need to use $HOTP$ algorithm ($HMAC$ based one time password) described in RFC 4226 [10] instead of $HMAC$ while calculating HPF values. RFC 4226 [10] which defines HOTP defines the function of data fields in HOTP as $HOTP(K, C, D)$ where our data field is equal to the concatenated hashed pels and C is the counter value. That way we have

$HPF = HOTP(K, C, (HP1 + HP2+ … +HPn))$. We didn't primarily suggest HOTP as a method because of the problem of counter synchronization, but we see that this would not be a problem for high security authentication systems regardless if $HOTP$ or $TOTP$ as per RFC 6238 [11] is implemented. Also, with $HOTP$ or $TOTP$ we lose the caching side of the server's enrolment phase (step 7), which would raise the load on the server for multiple users, but we don't see this as a problem since hash functions are efficient to compute from such small data sets as used in this specific problem.

### 5. Discussion

The presented method is quite simple; however, it significantly increases the security of a system. As shown, the search space in case of password cracking was increased with the use of additional information from the EEG. Our approach of using EEG signals for authentication is ready to use as is, because we use two EEG signals that are reliable, which can also be controlled by the user. This method does not need expensive equipment or EEG laboratory conditions. Authentication is fast, the search space for a potential attacker is significantly higher, and what the most prominent is that method allows to the user to install the mental state into the password itself. Therefore this approach is applicable in systems where it is important that users be able to control their mental state. Additionally, if a user is under stress or is forced to enter the password, it would be harder to cheat the system. Finally, we come to the question of ethics of the application of this method. Since the very beginning of this paper, we point out that such systems are unlikely become a reality in everyday public use, but in special cases where we require strong two factor authentication its application is expected.

## 6. Conclusion

In this research our aim was to supplement some deficiencies in the current research with EEG devices in the context of user authentication with our small contribution. We stated that the current study has the lack of applicability outside of laboratory conditions, with less expensive equipment and we set the goal to create a new authentication method that will be applicable to typical an EEG device currently available on the market. From the results of the research, it is evident that such inexpensive devices have little potential for a system such as pass-thought, but are good at resolving the meaning of alpha and beta waves to infer about the mental state of users. Specifically, in our case the device was giving accurate measurements of concentration and relaxation of users, so we applied those signal values as an additional parameter for the enhancement of passwords. Hence, we have significantly increased the search space with a reduced number of characters a password, since the main security is based on the shared secret key. With our method the user can have a shorter password, which is broken to a set of pels, and to each pel the value of concentration and/or relaxation is assigned by user. While user authentication takes place, user states his name and identification is performed. The algorithm verifies whether each pel is matched with user's mental state. Although this approach has drawbacks, it takes into account the mental state of users and thereby prevents or disables forcing users to authenticate. It also disables tired and careless users access to sensitive system parts. The disadvantage of this method is reflected in the fact that there is still a risk of shoulder surfing. But it is reduced, since the potential attacker must monitor simultaneously users' password and the matching mental state. Also, by using our customised algorithm we can implement one time passwords so replay attacks become extremely hard if not impossible. The direction of our further research will move towards the elimination of the need to enter a password. It should focus on reducing signal recording time considering the current lengths, it should not exceed more than 60 seconds. It would be interesting to find out the limit, when brain wave signals become useful versus the time when does user become annoyed by the length of the measurement process. Further research should also consider music as an instrument for stimulating brain activity and to find characteristic vector of each individual user, of course, using the general constraints we had - low budget equipment that is generally available in the market.